\pdfoutput=1
\documentclass[prl,twocolumn,showpacs,showkeys,superscriptaddress,amsmath,amssymb,floatfix]{revtex4-1}

\usepackage{epsfig}
\usepackage{rotating}
\usepackage{color}
\usepackage{bm}
\usepackage{hyperref}
\usepackage{mathtools}

\DeclarePairedDelimiter\ket{\lvert}{\rangle}
\DeclarePairedDelimiterX\braket[2]{\langle}{\rangle}{#1 \delimsize\vert #2}

\begin{document}


    \title{First-principles theory of proximity spin-orbit torque on a two-dimensional magnet: Current-driven antiferromagnet-to-ferromagnet reversible transition in bilayer CrI$_3$}
	
	\author{Kapildeb Dolui}
	\affiliation{Department of Physics and Astronomy, University of Delaware, Newark, DE 19716, USA}
	\author{Marko D. Petrovi\'{c}}
	\affiliation{Department of Physics and Astronomy, University of Delaware, Newark, DE 19716, USA}
	\affiliation{Department of Mathematical Sciences, University of Delaware, Newark, DE 19716, USA}
	\author{Klaus Zollner}
	\affiliation{Institute for Theoretical Physics, University of Regensburg, Regensburg 93040, Germany}
	\author{Petr Plech\'{a}\v{c}}
	\affiliation{Department of Mathematical Sciences, University of Delaware, Newark, DE 19716, USA}
	\author{Jaroslav Fabian}
	\affiliation{Institute for Theoretical Physics, University of Regensburg, Regensburg 93040, Germany}
	\author{Branislav K. Nikoli\'{c}}
	\email{bnikolic@udel.edu}
	\affiliation{Department of Physics and Astronomy, University of Delaware, Newark, DE 19716, USA}
	
	\begin{abstract}
		The recently discovered two-dimensional (2D) magnetic insulator CrI$_3$ is an intriguing case for basic research and spintronic applications since it is a ferromagnet in the bulk, but an antiferromagnet in bilayer form, with its magnetic ordering amenable to external manipulations. Using first-principles quantum transport approach, we predict that injecting unpolarized charge current parallel to the interface of bilayer-CrI$_3$/monolayer-TaSe$_2$ van der Waals heterostructure will induce spin-orbit torque (SOT) and thereby driven dynamics of magnetization on the first monolayer of CrI$_3$ in direct contact with TaSe$_2$. By combining calculated complex angular dependence of SOT with the Landau-Lifshitz-Gilbert equation for classical dynamics of magnetization, we demonstrate that current pulses can switch the direction of magnetization on the first monolayer to become parallel to that of the second monolayer, thereby converting CrI$_3$ from antiferromagnet to ferromagnet {\em while not requiring any external magnetic field}. We explain the mechanism of this reversible {\em current-driven nonequilibrium phase transition} by showing that first monolayer of CrI$_3$ carries current due to evanescent wavefunctions injected by metallic transition metal dichalcogenide TaSe$_2$, while concurrently acquiring strong spin-orbit coupling (SOC) via such proximity effect, whereas the second monolayer of CrI$_3$ remains insulating. The transition can be detected by passing vertical read current through the vdW heterostructure, encapsulated by bilayer of hexagonal boron nitride and sandwiched between graphite electrodes, where we find tunneling magnetoresistance of $\simeq 240$\%.
	\end{abstract}
	
	\maketitle
    {\em Introduction}.---The recent discovery of two-dimensional (2D) magnets derived from layered van der Waals (vdW) materials~\cite{Gong2017,Huang2017a} has opened new avenues for basic research on low-dimensional magnetism~\cite{Gibertini2019,Burch2018} and potential applications in spintronics~\cite{Cortie2019,Gong2019,Li2019,Alghamdi2019,Wang2019}. Their magnetic phases can substantially differ from those in conventional bulk magnetic materials due to large structural anisotropy which makes possible different sign and magnitude of intralayer $J_{\rm intra}$ and interlayer $J_{\rm inter}$ exchange coupling between localized magnetic moments. For example, $J_{\rm intra}$ and $J_{\rm inter}$ are ferromagnetic and antiferromagetic, respectively, between magnetic moments on Cr atoms within bilayer of CrI$_3$, which eventually becomes an antiferromagnetic insulator with N\'eel temperature \mbox{$T_\mathrm{N} \simeq 61$ K}~\cite{Huang2017a}. In such antiferromagnet spins have opposite orientation in the two monolayers, whereas monolayer, trilayer and bulk CrI$_3$ are ferromagnetic. Thus, bilayer of CrI$_3$ can also be viewed as two monolayer ferromagnets that are antiferromagnetically coupled to each other. The monolayer CrI$_3$ circumvents the Mermin-Wagner theorem~\cite{Mermin1966}, where thermal fluctuations destroy long-range magnetic order in 2D, by exhibiting strong uniaxial perpendicular magnetic anisotropy (PMA) which removes rotational invariance and effectively makes it a realization of the Ising model~\cite{Gibertini2019}. The PMA is also required for high-density device applications.

    \begin{figure}
    	\includegraphics[width=0.49\textwidth]{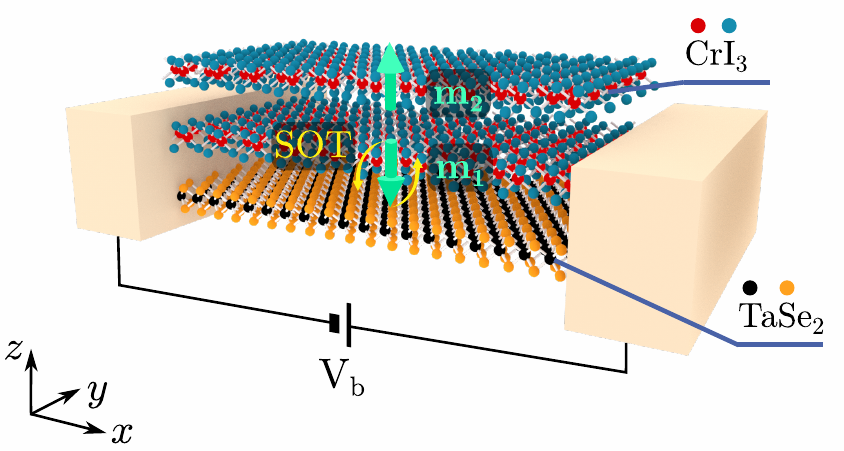}
    	\caption{Schematic view of CrI$_3$/TaSe$_2$ vdW heterostructure consisting of an insulating antiferromagnetic bilayer of CrI$_3$ and a nonmagnetic metallic monolayer TMD TaSe$_2$. The unpolarized charge current is injected parallel to the interface by a small applied bias voltage  V$_{\rm b}$ between the left and right macroscopic reservoirs. The current flows through monolayer of TaSe$_2$, as well as through first monolayer of CrI$_3$ due to evanescent wavefunctions injected into it by TaSe$_2$. The unit vectors of magnetizations on the two CrI$_3$ monolayers are denoted by $\mathbf{m}_1$ and $\mathbf{m}_2$. The heterostructure is assumed to be infinite in the $xy$-plane.}
    	\label{fig:device}
    \end{figure}

	The layer and stacking order~\cite{Sivadas2018,Jiang2019} dependence of electronic and spin structure as a new knob---together with possibilities for external manipulation via gating, straining and coupling to other 2D materials within vdW heterostructures---allows for dramatic changes of magnetic ordering of 2D magnets which is not possible with conventional bulk magnetic materials. For example, very recent experiments~\cite{Huang2018,Jiang2018,Jiang2018a} have demonstrated antiferromagnet (AFM) to ferromagnet (FM) phase transition in bilayer of CrI$_3$ by applying an external electric field via gate voltage or by electrostatic doping. While these effects offer potential building blocks~\cite{Locatelli2014} for an ultralow-dissipation nonvolatile memory, at present they employ cumbersome external magnetic fields which cannot be generated on nanoscale as required for integration with other elements of a circuit. Furthermore, reading the change of magnetic state within a circuit requires to pass a current through such devices~\cite{Zhou2019}, as demonstrated recently using an unconventional magnetic tunnel junctions where bilayer of  CrI$_3$ functions as the spin-filter sandwiched between metallic electrodes (such as graphene) with current flowing perpendicular to the interface~\cite{Song2018,Klein2018,Wang2018d,Song2019a}. 
	
	An alternative for magnetization switching is to inject a current through 2D  magnet and drive its magnetization dynamics via spin torque, as 
	exemplified by very recent experiments~\cite{Alghamdi2019,Wang2019} on spin-orbit torque (SOT)~\cite{Manchon2019,Ramaswamy2018} driven magnetization dynamics in Fe$_3$GeTe$_2$/Pt heterostructures. However, the layers employed in these experiments were much thicker than the ultimate limit envisaged using vdW heterostructures composed of just a few atomically thin layers. They are flat and ensure highly transparent interfaces, so that drastically smaller energy consumption per switching cycle can be achieved. These experiments have also relied on Fe$_3$GeTe$_2$ being a metallic vdW ferromagnet~\cite{Gibertini2019}, so that  CrI$_3$ bilayer with an energy gap is at first sight not suitable for SOT-operated devices.

	Here we employ first-principle quantum transport framework, which combines~\cite{Nikolic2018,Belashchenko2019,Belashchenko2019a} nonequilibrium Green functions (NEGFs)~\cite{Stefanucci2013} for two-terminal devices with noncollinear density functional theory (ncDFT) calculations~\cite{Capelle2001,Eich2013a},  to predict that the AFM-FM {\em nonequilibrium} phase transition can be induced by SOT in  bilayer-CrI$_3$/monolayer-TaSe$_2$ vdW lateral heterostructure depicted in Fig.~\ref{fig:device} where unpolarized charge current is injected parallel to the interface. The monolayer of metallic TaSe$_2$ is chosen in 1H-phase for which lattice mismatch between TaSe$_2$ and CrI$_3$ is as small as 0.1\%, while inversion symmetry of  TaSe$_2$ is broken to create large spin-orbit coupling (SOC). 

	\begin{figure}
		\center
		\includegraphics[scale=1.0]{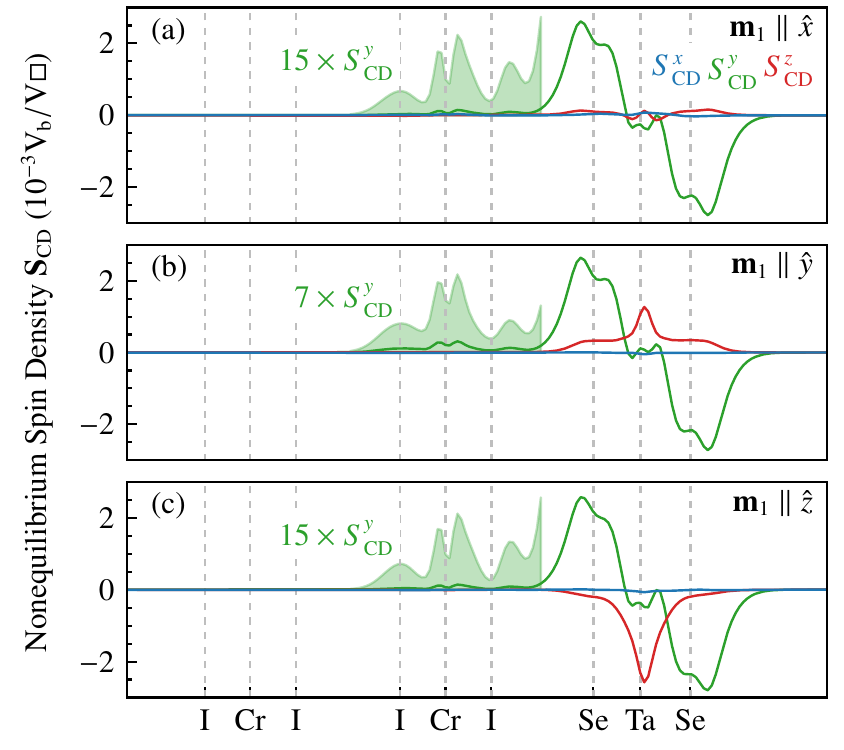}
		\caption{The current-driven nonequilibrium spin density $\mathbf{S}_{\rm CD}=(S_\mathrm{CD}^x,S_\mathrm{CD}^y,S_\mathrm{CD}^z)$ in the linear-response regime within bilayer-CrI$_3$/monolayer-TaSe$_2$ vdW heterostructure for: (a) ${\bf m}_{1} \parallel \hat{x}$; (b) ${\bf m}_{1} \parallel \hat{y}$; and  (c) ${\bf m}_{1} \parallel \hat{z}$. Vertical dashed lines indicate the position of each atomic plane. The area of the common rectangular supercell of vdW heterostructure in Fig.~\ref{fig:device} is denoted by $\Box = 2\sqrt{3}a^2$, where \mbox{$a$ = 6.85~\AA{}} is the lattice constant of bulk CrI$_3$. Shaded green areas represent rescaled $S_\mathrm{CD}^y$ in the spatial region  of the first monolayer of CrI$_3$ which is in direct contact with monolayer of TaSe$_2$.}
		\label{fig:spin_density}
	\end{figure}

\begin{figure}
    	\center
    	\includegraphics[scale=0.52]{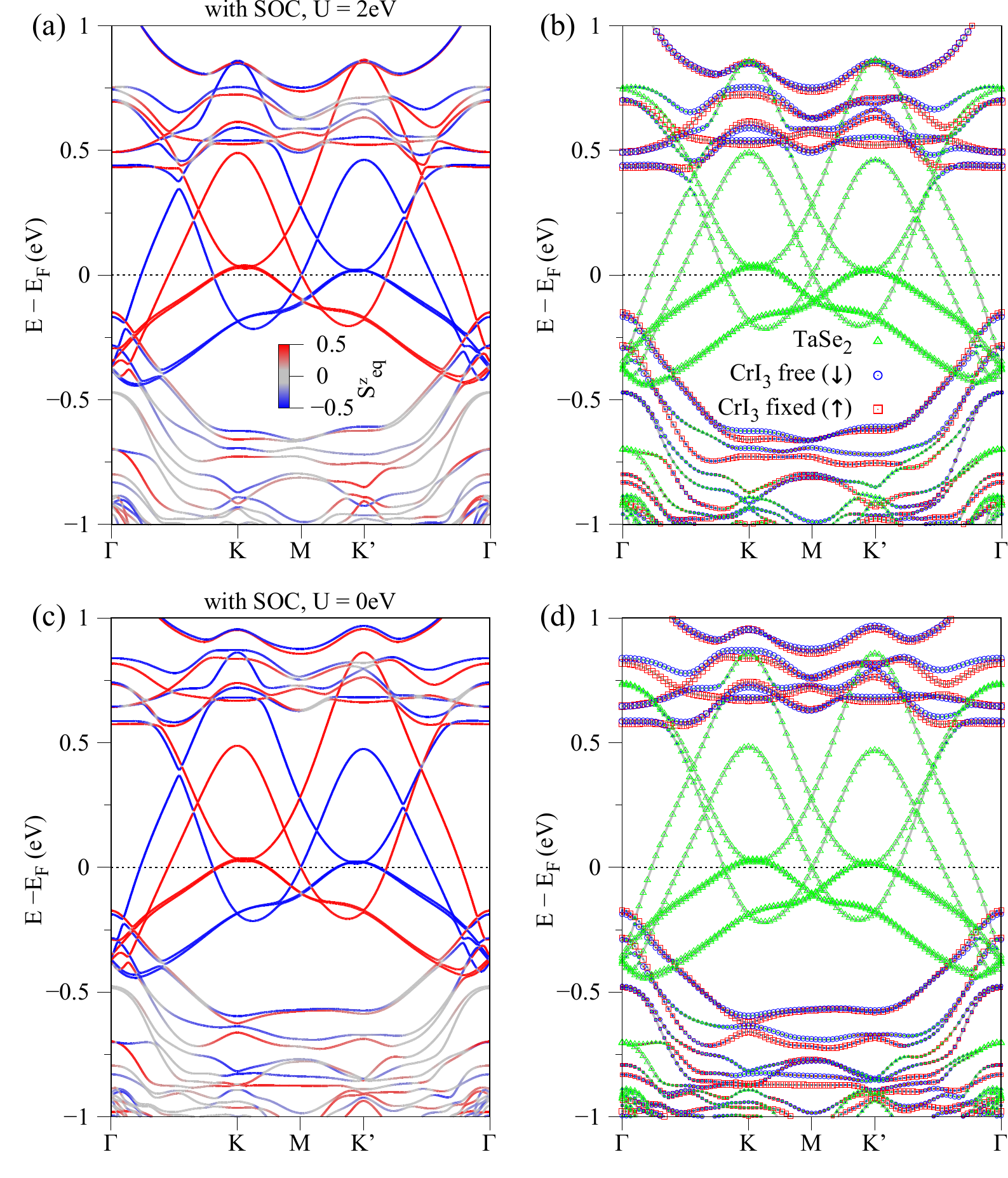}
    	\caption{First-principles-computed bands of bilayer-CrI$_3$/monolayer-TaSe$_2$ vdW heterostructure with 
    	SOC turned on. We use Hubbard $U = 2.0$~eV and $U = 0$~eV in panels (a) and (c), respectively, in ncDFT+U calculations where the color corresponds 
    	to spin expectation value in equilibrium, $S^z_\mathrm{eq} = \mathrm{Tr}\,[\boldsymbol{\rho}_\mathrm{eq} \hat{\sigma}_z]$. 
    	Panels (b) and (d) show the same band structures as (a) and (c), respectively, but with different colored symbols corresponding to projections onto different monolayers. } 
    	\label{fig:dftbands}
    \end{figure}

	Both conventional spin-transfer torque (in the absence of SOC and in geometries with two FM layers with noncollinear magnetizations~\cite{Nikolic2018,Ellis2017}) and SOT (in geometries with one FM layer but requiring interfacial or bulk SOC effects~\cite{Nikolic2018,Belashchenko2019,Belashchenko2019a,Freimuth2014,Mahfouzi2018}) can be described microscopically and independently of particular physical mechanism~\cite{Manchon2019,Belashchenko2019a} as a consequence of the interaction between current-driven (CD) nonequilibrium spin density~\cite{Edelstein1990,Aronov1989,Chang2015} of conduction electrons $\mathbf{S}_{\rm CD}(\mathbf{r})$ and a nonzero exchange-correlation (XC) magnetic field $\mathbf{B}_\mathrm{XC}(\mathbf{r})$~\cite{Capelle2001,Eich2013a} present in equilibrium. Their cross product, $\mathbf{S}_\mathrm{CD}(\mathbf{r}) \times \mathbf{B}_\mathrm{XC}(\mathbf{r})$, is local torque at some point in space $\mathbf{r}$, so that total torque is obtained by integration~\cite{Nikolic2018,Belashchenko2019,Ellis2017} 
	\begin{equation}\label{eq:sotrealspace}
	\mathbf{T} = \int \! d^3r \, \mathbf{S}_\mathrm{CD}(\mathbf{r}) \times \mathbf{B}_\mathrm{XC}(\mathbf{r}).
	\end{equation}
	While $\mathbf{B}_\mathrm{XC}(\mathbf{r})$ is nonzero in both monolayer and bilayer of CrI$_3$ due to long-range magnetic ordering, $\mathbf{S}_{\rm CD}(\mathbf{r})$ appears {\em only} on the monolayer of CrI$_3$ that is in direct contact with monolayer of TaSe$_2$, as demonstrated by Fig.~\ref{fig:spin_density}. This is due to the proximity effect where evanescent wavefunctions from TaSe$_2$ penetrate [Figs.~\ref{fig:spin_density} and ~\ref{fig:dftbands}] up to the first monolayer of CrI$_3$ to make it a current carrier,  while also bringing~\cite{Marmolejo-Tejada2017} SOC from TaSe$_2$ to ensure that $\mathbf{S}_{\rm CD}(\mathbf{r})$ is not collinear to $\mathbf{B}_\mathrm{XC}(\mathbf{r})$. The giant SOC hosted by TaSe$_2$ itself due to inversion symmetry breaking in ultrathin layers of transition metal dichalcogenides (TMDs)~\cite{Zhu2011a,Ge2012} is confirmed by large $\mathbf{S}_{\rm CD}(\mathbf{r})$ within the spatial region of TaSe$_2$ monolayer in Fig.~\ref{fig:spin_density}. 
	
\begin{figure}
		\includegraphics[scale=1.0]{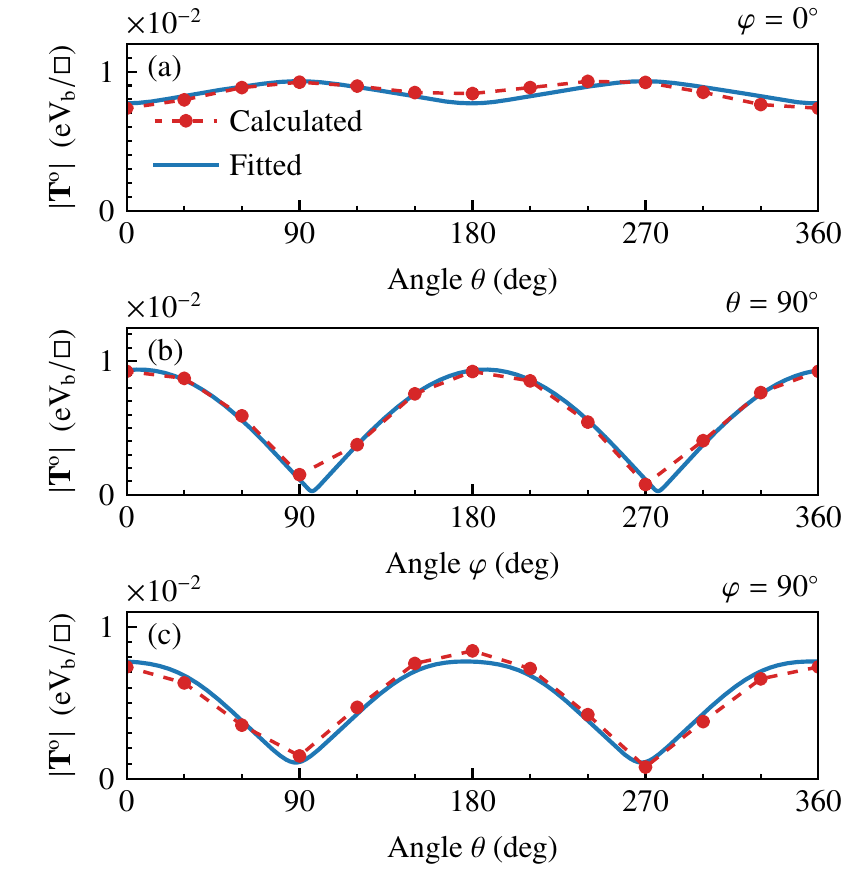}
		\caption{Azimuthal ($\theta$) and polar ($\phi$) angle dependence of the magnitude of SOT component $|\mathbf{T}^{\rm o}|$ (odd in $\mathbf{m}_1$)  for different orientation of the magnetization \mbox{$\mathbf{m}_1 = (\sin\theta \cos\phi, \sin \theta \sin \phi, cos~\theta)$} on the first monolayer of CrI$_3$. The magnetization is rotated within the (a) $xz$-plane; (b) $xy$-plane; and (c) $yz$-plane. Solid blue lines are fit to NEGF+ncDFT-computed SOT values (red dots) using Eq.~\eqref{eq:sotexpansion} with fitting parameters from Table~\ref{tab:torque}.}
		\label{fig:torque}
	\end{figure}

\begin{figure}
		\includegraphics[scale=1.0]{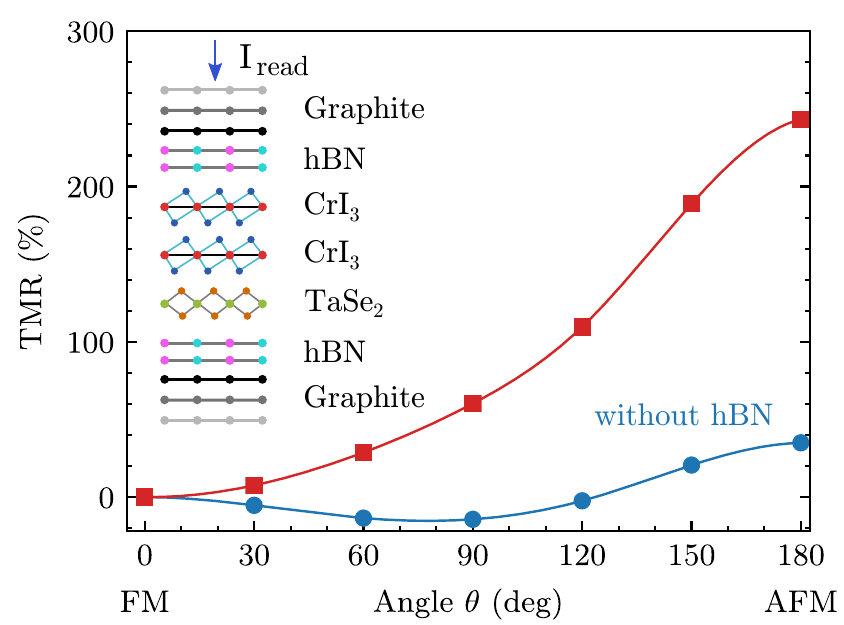}
		\caption{TMR vs. angle $\theta$  between magnetizations $\mathbf{m}_1$ and $\mathbf{m}_2$ on two monolayers of CrI$_3$ in Fig.~\ref{fig:device} for vertical read current~\cite{Zhou2019} flowing perpendicularly (i.e., along the $z$-axis in Fig.~\ref{fig:device}) through bilayer-CrI$_3$/monolayer-TaSe$_2$ vdW heterostructure. The heterostructure is sandwiched between two semi-infinite graphite leads with (red squares) and without (blue circles) bilayer of hBN inserted between the leads and the vdW heterostructure, as illustrated in the inset.}
		\label{fig:tmr}
	\end{figure}

	The SOT vector in Eq.~\eqref{eq:sotrealspace}, with its complex angular dependence [Fig.~\ref{fig:torque}] on the direction of magnetization (along the unit vector $\mathbf{m}_1$ in Fig.~\ref{fig:device}) of the first monolayer of CrI$_3$, is combined in a multiscale fashion~\cite{Ellis2017,Petrovic2018} with the classical dynamics of magnetization  governed by the Landau-Lifshitz-Gilbert  (LLG) equation to demonstrate reversible switching of $\mathbf{m}_1$ [Fig.~\ref{fig:llg}(b) and the accompanying movie in the Supplemental Material (SM)~\cite{sm}] from -$z$ to $+z$ direction by current pulses and, thereby, transition from AFM to FM phase of CrI$_3$ bilayer. The dynamics of $\mathbf{m}_1$ can be detected by passing vertical read current along the $z$-axis~\cite{Zhou2019}, where we compute the tunneling magnetoresistance (TMR) of $\simeq 240$\% [Fig.~\ref{fig:tmr}] due to AFM-FM transition of CrI$_3$ bilayer. For such as a scheme, we assume that bilayer-CrI$_3$/monolayer-TaSe$_2$ vdW heterostructure is sandwiched between two metallic semi-infinite graphite electrodes along the $z$-axis with hexagonal BN (hBN) bilayers inserted between the leads and vdW heterostructure [inset of Fig.~\ref{fig:tmr}].

	{\em Methodology}.---We employ the interface builder in {\tt QuantumATK}~\cite{quantumatk}  package to construct a unit cell of vdW heterostructure in Fig.~\ref{fig:device} while starting from experimental lattice constants of CrI$_3$ and  TaSe$_2$ layers. In order to determine the interlayer distance 
	between CrI$_3$ and TaSe$_2$, we perform DFT calculations with Perdew-Burke-Ernzerhof (PBE) parametrization of the  generalized gradient approximation (GGA) for the XC functional, including Grimme D2~\cite{Grimme2006} vdW corrections.

	\begin{figure}
		\includegraphics[scale=1.0]{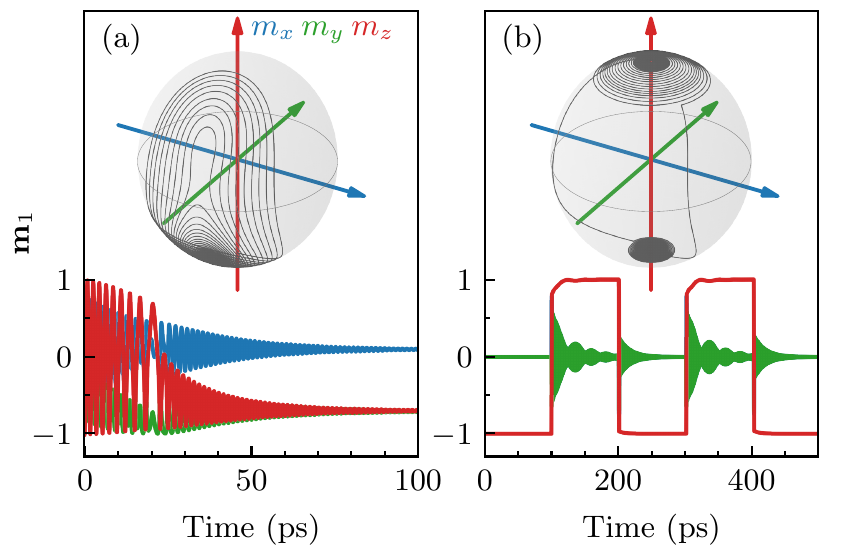}
		\caption{Classical dynamics of magnetization $\mathbf{m}_1(t)$ on the first monolayer of CrI$_3$ which is exchange coupled [Eq.~\eqref{eq:Hc}] to magnetization $\mathbf{m}_2$ on the second monolayer of CrI$_3$ while experiencing SOT from Fig.~\ref{fig:torque}. The dynamics is obtained by solving two coupled LLG equations [Eq.~\eqref{eq:llg}] for $\mathbf{m}_1(t)$ and $\mathbf{m}_2(t)$, where the later remains nearly fixed ($m_2^z \approx 1$; while $m_2^x$ and $m_2^y$ perform small oscillations around zero). The bias voltage $V_{\rm b} = 0.2$~V is dc in (a), while in (b) we use  a sequence of short rectangular voltage pulses of the same amplitude with pulse duration \mbox{$\delta  t_{\rm ON} = 0.68$~ps} followed by a pause of \mbox{$\delta t_{\rm OFF} = 100$~ps} during which no voltage is applied. A movie animating $\mathbf{m}_1(t)$ in panel (b), as well as $\mathbf{m}_2(t)$, is provided in the SM~\cite{sm}.} 
		\label{fig:llg}
	\end{figure}

	In addition, we employ  ncDFT+U calculations using {\tt Quantum ESPRESSO}~\cite{Giannozzi2009} package to examine how nonzero Hubbard $U$~\cite{Liechtenstein1995} affects the bands of the vdW heterostructures since nonzero $U$ has been utilized before~\cite{Lado2017,Zollner2019} as  
	a cure for the band gap problem in CrI$_3$~\cite{Zhang2015}. For this purpose, we use PBE parametrization of GGA for the XC functional; fully relativistic pseudopotentials with the projector augmented wave method~\cite{Kresse1999} for describing electron-core interactions; energy cutoff of $550$~Ry for plane wave basis set; and  $k$-point sampling of $30\times 30\times 1$ for self-consistent calculations. Comparing the cases with $U = 0$ and $U = 2$~eV in Fig.~\ref{fig:dftbands} shows that $U \neq 0$ barely changes the band structure within the energy window $\pm 1$~eV, with only the conduction band of CrI$_3$ experiencing a shift in energy of about $\simeq 0.15$~eV [Fig.~\ref{fig:dftbands}(b),(d)] at the $\Gamma$ point and energies around 0.5~eV. This is because Hubbard $U$ acts~\cite{Lado2017,Zollner2019} on $d$-orbitals of Cr whose bands are higher in energy. States near the Fermi level \mbox{$E-E_F=0$}, which are responsible for charge and spin transport properties in the linear-response transport regime, are formed by  TaSe$_2$ monolayer [Fig.~\ref{fig:dftbands}(b),(d)]. Due to short range of the proximity effect, only the bands of the first monolayer of  CrI$_3$ hybridize with the bands of TaSe$_2$, such as spin-down bands near the K valley at energies 0.5--0.8~eV [Fig.~\ref{fig:dftbands}(a),(b)].

	The NEGF+ncDFT formalism~\cite{Nikolic2018,Belashchenko2019,Ellis2017}, which combines self-consistent Hamiltonian from ncDFT calculations (with $U=0$ based on Fig.~\ref{fig:dftbands}) with nonequilibrium density matrix and current calculations from NEGF calculations, makes it possible to compute spin torque in arbitrary device geometry at small or finite bias voltage. The single-particle Kohn-Sham (KS) Hamiltonian in ncDFT is given by
	\begin{equation}\label{eq:ham_ks}
	\hat{H}_\mathrm{KS}=-\frac{\hbar^2\nabla^2}{2m} + V_\mathrm{ext}(\mathbf{r})  + V_\mathrm{H}(\mathbf{r}) + V_\mathrm{XC}(\mathbf{r}) -  \boldsymbol{\sigma}\cdot\mathbf{B}_\mathrm{XC}(\mathbf{r}),
	\end{equation}
	where $\boldsymbol{\sigma} = (\hat{\sigma}_x, \hat{\sigma}_y,\hat{\sigma}_z)$ is the vector of the Pauli matrices;  $V_\mathrm{ext}(\mathbf{r})$, $V_\mathrm{H}(\mathbf{r})$ and  \mbox{$V_\mathrm{XC}(\mathbf{r}) = \delta E_\mathrm{XC}[n(\mathbf{r}),{\bm \mu}(\mathbf{r})]/\delta n(\mathbf{r})$} are the external, Hartree and XC potentials, respectively; and the XC magnetic field, \mbox{$\mathbf{B}_\mathrm{XC}(\mathbf{r}) = \delta E_\mathrm{XC}[n(\mathbf{r}),{\bm \mu}(\mathbf{r})]/\delta {\bm \mu}(\mathbf{r})$}, is functional derivative with respect to the vector  magnetization density ${\bm \mu}(\mathbf{r})$. The extension of DFT to the case of spin-polarized systems is formally derived in terms of ${\bm \mu}(\mathbf{r})$ and total electron density $n(\mathbf{r})$, where in collinear DFT ${\bm \mu}(\mathbf{r})$ points in the same direction at all points in space, while in ncDFT ${\bm \mu}(\mathbf{r})$ can point in an arbitrary direction~\cite{Capelle2001,Eich2013a}.

	The heterostructure in Fig.~\ref{fig:device} is split into the central region and left (L) and right (R) semi-infinite leads, all of which are composed of the same CrI$_3$/TaSe$_2$ trilayer. The self-energies of the leads $\boldsymbol{\Sigma}_\mathrm{L,R}(E)$ and the Hamiltonian $\hat{H}_\mathrm{KS}$ of the central region are obtained from ncDFT calculations within {\tt QuantumATK} package~\cite{quantumatk} using: PBE parametrization of GGA for the XC functional; norm-conserving fully relativistic pseudopotentials of the type SG15-SO~\cite{quantumatk,Schlipf2015} for describing electron-core interactions; and SG15 (medium) numerical linear combination of atomic orbitals (LCAO)  basis set~\cite{Schlipf2015}. Periodic boundary conditions are employed in the plane perpendicular to the transport direction with grids of 1$\times$101~$k$-point (lateral device setup in Fig.~\ref{fig:device} for SOT calculations) and 25$\times$25 $k$-point (vertical device setup in the inset of Fig.~\ref{fig:tmr} for TMR calculations). The energy mesh cutoff for the real-space grid is chosen as 100 Hartree.
    
	The lesser Green function (GF), \mbox{$\mathbf{G}^{<}(E) = i \mathbf{G}(E)[f_L(E)\boldsymbol{\Gamma}_L(E)+f_R(E)\boldsymbol{\Gamma}_R(E)]\bold{G}^{\dagger}(E)$}, of NEGF formalism makes it possible to construct the nonequilibrium density matrix~\cite{Stefanucci2013}
	\begin{equation}\label{eq:noneqrho}
	\boldsymbol{\rho}_{\rm neq} = \frac{1}{2\pi i} \int_{-\infty}^{\infty}\!\! dE\, \mathbf{G}^{<}(E),
	\end{equation}
	in the steady-state and elastic transport regime. Here \mbox{$\mathbf{G} =  [E \bold{\Lambda} - \mathbf{H}_\mathrm{KS}-\boldsymbol{\Sigma}_\mathrm{L}(E,V_\mathrm{L})-\boldsymbol{\Sigma}_\mathrm{R}(E,V_\mathrm{R})]^{-1}$} is the retarded GF, \mbox{$f_\mathrm{L,R}(E)=f(E-eV_\mathrm{L,R})$} are the shifted Fermi functions of the macroscopic reservoirs into which semi-infinite leads terminate; $V_\mathrm{b}=V_\mathrm{L} - V_\mathrm{R}$ is the applied bias voltage between them; and $\boldsymbol{\Gamma}_\mathrm{L,R}(E) = i[\boldsymbol{\Sigma}_\mathrm{L,R}(E)- \boldsymbol{\Sigma}^{\dagger}_\mathrm{L,R}(E)]$ are the level broadening matrices.  For lateral heterostructure [Fig.~\ref{fig:device}], all matrices---$\mathbf{H}_\mathrm{KS}$, $\mathbf{G}$, $\mathbf{G}^{<}$, $\boldsymbol{\Gamma}_\mathrm{L,R}$ and $\boldsymbol{\rho}_{\rm neq}$---depend on $k_y$, while for vertical heterostructure [inset of Fig.~\ref{fig:tmr}] they depend on $(k_x,k_y)$. Due to nonorthogonality of LCAO basis set $\ket{\phi_n}$, we also use the overlap matrix ${\bm \Lambda}$ composed of elements $\braket{\phi_i}{\phi_j}$. 
	
	The CD part of the nonequilibrium density matrix, $\boldsymbol{\rho}_\mathrm{CD}(k_y) = \boldsymbol{\rho}_\mathrm{neq}(k_y) -\boldsymbol{\rho}_\mathrm{eq}(k_y)$, is obtained by subtracting the equilibrium density matrix $\boldsymbol{\rho}_\mathrm{eq}(k_y)$ for $V_\mathrm{L}=V_\mathrm{R}$. This yields 
	\begin{equation}\label{eq:scd}
	\mathbf{S}_{\rm CD}(k_y) = \mathrm{Tr}[\boldsymbol{\rho}_{\rm CD}(k_y)\boldsymbol{\sigma} \boldsymbol{\Lambda}^{-1}],
	\end{equation}
	and SOT
	\begin{equation}\label{eq:sotvector}
	\mathbf{T} = \frac {1}{\Omega_\mathrm{BZ}} \int_\mathrm{BZ}\! dk_y\, [\boldsymbol{\mathrm{S}}_{\rm CD}(k_y) \times \mathbf{B}_\mathrm{XC}(k_y)],
	\end{equation}
	which we compute by performing trace in the LCAO basis [instead of in real space as in Eq.~\eqref{eq:sotrealspace}], and additional integration over the one-dimensional Brillouin zone (BZ) of length $\Omega_\mathrm{BZ}$ is performed.
	
	{\em Nonequilibrium spin density}.---The $xy$-plane averaged $\mathbf{S}_{\mathrm{CD}}$ is plotted in Fig.~\ref{fig:spin_density} for three representative orientations of the magnetization $\mathbf{m}_1 \parallel \{\hat x,\hat y,\hat z\}$ on the first monolayer of CrI$_3$. The nonequilibrium spin density is zero on the second monolayer of CrI$_3$, which confirms that evanescent wavefunctions originating from metallic TaSe$_2$ monolayer and the {\em spin-orbit proximity effect} carried by them decay exponentially fast, so they are able to reach only the first monolayer of CrI$_3$. Independently of the  orientation of $\mathbf{m}_1$, the magnitude of $\mathbf{S}_{\mathrm{CD}}$ within the first monolayer of CrI$_3$ is mainly dominated by its $y$-component, which is an order of magnitude smaller than $\mathbf{S}_{\mathrm{CD}}$ within TaSe$_2$. Concurrently, magnetic proximity effect from CrI$_3$ induces small magnetization into the monolayer of TaSe$_2$ with magnetic moments on Ta and Se atoms being \mbox{$0.008$ $\mu_{\mathrm{B}}$} and \mbox{0.001~$\mu_{\mathrm{B}}$}, respectively, where $\mu_\mathrm{B}$ is the Bohr magneton. In comparison, we compute magnetic moments on Cr and I atoms as \mbox{$\mu_\mathrm{Cr} = 3.43$~$\mu_\mathrm{B}$} and \mbox{$\mu_\mathrm{I} = 0.14$ $\mu_\mathrm{B}$}, respectively. The component $S_{\mathrm{CD}}^y$ within TaSe$_2$ monolayer is insensitive to $\mathbf{m}_1$, while S$^{x}_{\mathrm{CD}}$ remains negligible. Unlike the surface of topological insulator~\cite{Chang2015} or heavy metals, where $\mathbf{S}_{\mathrm{CD}}$ is confined to the plane in accord with the phenomenology of the standard inverse spin-galvanic (or Edelstein)  effect~\cite{Edelstein1990,Aronov1989}, TaSe$_2$ can exhibit large out-of-plane component S$^{z}_{\mathrm{CD}}$ which is sensitive to the orientation of $\mathbf{m}_1$ and it is highly sought for SOT-operated device applications~\cite{MacNeill2017}.
		
	\textit{Angular dependence of SOT}.---The SOT vector  can be decomposed~\cite{Freimuth2014,Garello2013}, \mbox{$\mathbf{T} = \mathbf{T}^\mathrm{e} + \mathbf{T}^\mathrm{o}$}, into odd (o) and even (e) components with respect to the magnetization $\mathbf{m}_1$. They can be computed directly   
	from Eqs.~\eqref{eq:scd} and ~\eqref{eq:sotvector} by using the respective components of the nonequilibrium density matrix, $\boldsymbol{\rho}_\mathrm{CD} = \boldsymbol{\rho}_\mathrm{CD}^\mathrm{e} + \boldsymbol{\rho}_\mathrm{CD}^\mathrm{o}$, as introduced in Refs.~\cite{Nikolic2018,Mahfouzi2016}. Due to the absence of the bulk in the case of monolayer TaSe$_2$, the vdW heterostructure in Fig.~\ref{fig:device} does not generate vertical spin Hall current along the $z$-axis as one of the mechanisms for $\mathbf{T}^\mathrm{e}$. Other interfacially based mechanisms~\cite{Belashchenko2019a} for $\mathbf{T}^\mathrm{e} \neq 0$ require backscattering of electrons~\cite{Nikolic2018,Pesin2012a,Kalitsov2017,Zollner2019a}, which is absent in the ballistic transport regime we assume, so we find  $\mathbf{T}^\mathrm{e} \rightarrow 0$. 
	
	The nonzero  $\mathbf{T}^{\rm o}$ component, computed from NEGF+ncDFT formalism as dots in Fig.~\ref{fig:torque}, can be fitted by a function defined as an infinite series~\cite{Garello2013}
	\begin{eqnarray}\label{eq:sotexpansion}
	{\bf{T}^{\rm o}} & = & (\mathbf{p} \times \mathbf{m}_1) 
	\Big[\sum_{n=0}^{\infty} \tau_{n\alpha}^\mathrm{o} 
	|\hat{z} \times \mathbf{m}_1|^{2n}\Big]
	\nonumber \\
	& & +\mathbf{m}_1 \times 
	(\hat {z} \times \mathbf{m}_1)
	(\mathbf{m}_1 \cdot \hat{x})
	\Big[\sum_{n=0}^{\infty} \tau_{n\beta}^\mathrm{o}
	|{\hat z} \times \mathbf{m}_1|^{2n}\Big],
	\end{eqnarray}
	assuming that current flows along the $x$-axis as in Fig.~\ref{fig:device}. Note that other expansions, such as in terms of orthonormal vector spherical harmonics, can also be employed to define the fitting function~\cite{Belashchenko2019a}. Here $\tau^\mathrm{o}_{n\alpha}$ and $\tau^\mathrm{o}_{n\beta}$ are the fitting parameters and $\mathbf{p}$ is the unit vector along the reference direction set by  current-induced nonequilibrium spin density, such that $\mathbf{T}^{\rm o}=0$ when $\mathbf{m}_1 \parallel \mathbf{p}$. In simple systems, like the Rashba spin-split 2D electron gas~\cite{Edelstein1990} or metallic surface of topological insulator~\cite{Chang2015} in contact with FM layer,  $\mathbf{p} \parallel \hat{y}$ (assuming injected current along $\hat{x}$) is determined by symmetry arguments~\cite{Belashchenko2019}. However, for more complicated systems it has to be calculated, and we find $\mathbf{p} \equiv (\theta=88^\circ,\phi=98^\circ)$ instead of often na\"{i}vely assumed $\mathbf{p} \equiv (\theta=90^\circ,\phi=90^\circ) \parallel \hat{y}$. The lowest order term  $\tau^\mathrm{o}_{0\alpha}(\mathbf{p} \times \mathbf{m}_1)$ in Eq.~\eqref{eq:sotexpansion} is conventional  field-like torque~\cite{Manchon2019}, while higher terms can have properties of both field-like and damping-like torque~\cite{Belashchenko2019a} [the lowest order term $\tau^\mathrm{e}_{0\alpha} \mathbf{m}_1 \times (\mathbf{p} \times \mathbf{m}_1)$ in the expansion of $\mathbf{T}^\mathrm{e}$ is conventional damping-like torque~\cite{Garello2013}]. The value of $\tau^\mathrm{o}_{0\alpha}$, together with other non-negligible parameters in Eq.~\eqref{eq:sotexpansion}, is given in Table~\ref{tab:torque}.

	\begin{table}[t]
		\centering
		\begin{tabular}{c|c c c c | c c}
			\hline \hline
			$\mathbf{p} (\theta,\phi)$ & $\tau_{0\alpha}^\mathrm{o}$ 
			& $\tau_{1\alpha}^\mathrm{o}$ & $\tau_{2\alpha}^\mathrm{o}$ 
			& $\tau_{3\alpha}^\mathrm{o}$ & $\tau_{0\beta}^\mathrm{o}$ 
			& $\tau_{1\beta}^\mathrm{o}$ \\
			\hline 
			(88$^{\circ}$, 98$^{\circ}$)
			& 77.22 & 17.32 & -30.32 
			& 13.54 & -9.19 & -6.88 \\
			\hline \hline
		\end{tabular}
		\caption{The non-negligible coefficients (in units of $10^{-4}~\mathrm{eV_\mathrm{b}/\Box}$) in the expansion of ${\bf{T}^{\rm o}}$ in Eq.~\eqref{eq:sotexpansion} are obtained by fitting (solid lines) NEGF+ncDFT-computed angular dependence of SOT (dots) in Fig.~\ref{fig:torque} for  bilayer-CrI$_3$/monolayer-TaSe$_2$ vdW heterostructure.}
		\label{tab:torque}
	\end{table}

	\textit{SOT-driven classical dynamics of magnetization}.---The effective anisotropic classical Heisenberg model~\cite{Gibertini2019} for magnetic moments $\mathbf{m}_1$ and $\mathbf{m}_2$ on Cr atoms within two monolayers of CrI$_3$ in Fig.~\ref{fig:device} is given by
	\begin{equation}\label{eq:Hc}
	\mathcal{H} = -J_{12}(\mathbf{m}_1 \cdot \mathbf{m}_2) - \sum_{i=1,2} A_{\perp} ({m}_{i}^{z})^2,
	\end{equation}
	where the values for $J_{12}=-0.05$~meV, as the interlayer AFM exchange coupling, and $A_{\perp}= 2.0$~meV, as the PMA constant in the presence of TaSe$_2$ monolayer, are extracted from ncDFT calculations. They are close to the corresponding values obtained for isolated CrI$_3$ bilayer in previous ncDFT calculations~\cite{Zhang2015c,Sivadas2018}. 
	
	We simulate the classical dynamics $\mathbf{m}_1(t)$ by solving the LLG equation
	\begin{equation}\label{eq:llg}
	\frac{d \mathbf{m}_1}{dt} = -\gamma \mathbf{m}_1 \times \mathbf{B}^{\rm eff}_{1} + \lambda_1 \mathbf{m}_1 \times \frac{d\mathbf{m}_1}{dt} + \frac{\gamma}{\mu_\mathrm{Cr}} \mathbf{T}^\mathrm{o},
	\end{equation}
	where $\gamma$ is the gyromagnetic ratio; $\mathbf{B}^{\rm eff}_{1} = - \frac{1}{\mu_\mathrm{Cr}} \partial \mathcal{H} /\partial \mathbf{m}_{1}$ is the effective magnetic field due to interactions in the Hamiltonian in Eq.~\eqref{eq:Hc}; $\lambda_1$ is the Gilbert damping parameter; and SOT at arbitrary direction of $\mathbf{m}_1$ is given by Eq.~\eqref{eq:sotexpansion} with parameters in Table~\ref{tab:torque}. The LLG equation for $\mathbf{m}_2$ is the same as Eq.~\eqref{eq:llg}, but with $\mathbf{B}^{\rm eff}_{2}= - \frac{1}{\mu_\mathrm{Cr}} \partial \mathcal{H} /\partial \mathbf{m}_{2}$ and $\mathbf{T}^\mathrm{o} \equiv 0$ because no current flows through the second monolayer of CrI$_3$.
	
	The computed trajectories $\mathbf{m}_1(t)$ are plotted in Fig.~\ref{fig:llg}(a) for dc bias voltage, as well as for rectangular voltage pulses in Fig.~\ref{fig:llg}(b). The trajectories $\mathbf{m}_2(t)$ are trivial---$m_2^z(t) \approx 1$ while $m_2^x(t)$ and $m_2^y(t)$ perform small oscillations around zero---so they are not plotted. The time evolutions $\mathbf{m}_1(t)$ and  $\mathbf{m}_2(t)$ are also animated in the movie provided as the SM~\cite{sm}. For unpolarized charge current injected by dc bias, magnetization $\mathbf{m}_1$ switches from being antiparallel to $\mathbf{m}_2$ to a noncollinear direction within the $yz$-plane. The nonequilibrium and noncollinear configuration of $\mathbf{m}_1$ and $\mathbf{m}_2$  will return to AFM phase when $V_\mathrm{b}$ is turned off and the system goes back to equilibrium. On the other hand, using voltage pulses leads to AFM-FM transition with reversal from $\mathbf{m}_1 \parallel -\hat{z}$ to $\mathbf{m}_1 \parallel +\hat{z}$ while magnetization of the second layer remains $\mathbf{m}_2 \parallel +\hat{z}$. Such current-induced FM phase is stable in-between two pulses, on the proviso that $A_{\perp} > J_{12}$ in Eq.~\eqref{eq:Hc}, and can be {\em reversed} back to the AFM phase by the next pulse [Fig.~\ref{fig:llg}(b) and movie in the SM~\cite{sm}]. We assume different Gilbert damping parameters $\lambda_1 = 0.01 > \lambda_2 = 0.0001$ on two monolayers of CrI$_3$ due to the presence of TaSe$_2$ monolayer, but the actual value  $\lambda_1$ on the first monolayer of CrI$_3$  is likely smaller. Thus, we anticipate that the time needed to stabilize the FM phase would be of the order of \mbox{$\sim 1$~ns}, instead of \mbox{$\sim 100$~ps} in Fig.~\ref{fig:llg} (where $\lambda_1$ was tuned for such numerical convenience).  

	Since  $\mathbf{T}^\mathrm{o}$ forces precession of magnetization around the axis defined by $\mathbf{p}$, magnetization reversal in the voltage pulse setup is achieved by fine tuning the pulse duration $\delta t_{\rm ON}$ to half of the period of that precession. The Gilbert damping term, $\lambda_1  \mathbf{m}_1 \times d\mathbf{m}_1/dt$, does not play a role in this type of switching, although $\lambda_1$ together with PMA constant is critical to stabilize  the FM phase after the pulse is switched off, as shown in the movie in the SM~\cite{sm}. For instance,  when Gilbert damping  is set to zero, the magnetizations in both monolayers never fully align with the $z$-axis and instead continue to precess around it which renders the FM phase unstable. Note that more detailed LLG simulations would require to simulate more than two magnetic moments and their inhomogeneous switching in a particular device geometry, as often observed experimentally in ferromagnet/heavy-metal heterostructures~\cite{Baumgartner2017}, but our two-terminal device is homogeneous and translationally invariant within the $xy$-plane in Fig.~\ref{fig:device}. Also, in the presence of disorder and thereby induced voltage drop across the central region~\cite{Belashchenko2019,Belashchenko2019a,Kalitsov2017} we expect that $\mathbf{T}^\mathrm{e}$ would become nonzero and contribute to switching. 
		
	\textit{TMR as a  probe of AFM-FM transition}.---Finally, akin to experiments~\cite{Zhou2019} where SOT-driven magnetization switching has been probed by passing additional vertical read current through SOT devices operated by lateral current, we investigate angular dependence of TMR for vertical current  assumed to be injected between semi-infinite graphite leads along the $z$-axis sandwiching bilayer-CrI$_3$/monolayer-TaSe$_2$ [see inset in  Fig.~\ref{fig:tmr} for illustration].  We define angular dependence of TMR as  $\mathrm{TMR}(\theta)=[R(\theta)-R(0)]/R(0)$, where $R(0)$ is the resistance of FM phase with $\mathbf{m}_1 \parallel \hat z \parallel \mathbf{m}_2$ and  $R(\theta)$ is the resistance for angle $\theta$ between them. Thus, $R(\theta=180^\circ)$ corresponds to AFM phase. Note that $\mathrm{TMR}(\theta=180^\circ)$ recovers the conventional definition of TMR using only parallel and antiparallel configuration of magnetizations. In Fig.~\ref{fig:tmr} we obtain $\mathrm{TMR}(\theta=180^\circ) \simeq 240$\% when using additional hBN bilayers inserted between graphite leads and the vdW heterostructure. When hBN is removed, TMR drops to $\mathrm{TMR}(\theta=180^\circ) \simeq 40$\%, while exhibiting peculiar change of sign for angles between  $\theta=0^\circ$ and $\theta=180^\circ$ in accord with experimental observation reported in Ref.~\cite{Song2018} of few-layer-graphene/bilayer-CrI$_3$/few-layer-graphene junctions.
	
	
	\section{Acknowledgments}
	K.~D and B.~K.~N. were supported by DOE Grant  No.~DE-SC0016380. M.~D.~P. and P.~P. were supported by ARO MURI Award No.~W911NF-14-0247. K.~Z. and J.~F. were supported by DFG SPP~1666, SFB~1277. The supercomputing time was provided  by XSEDE, which is supported by NSF Grant No.~ACI-1053575.


\end{document}